\documentclass{article}

\usepackage{arxiv}

\usepackage[english]{babel}

\usepackage{amsmath}
\usepackage{amsfonts}
\usepackage{mathtools}
\usepackage{xcolor}
\usepackage{caption}
\usepackage{xurl}
\usepackage{array}
\usepackage{tabularx}
\usepackage{ragged2e}
\usepackage{graphicx}
\usepackage{cite}
\usepackage[colorlinks=true, allcolors=blue]{hyperref}
\usepackage{subcaption}
\usepackage{authblk}

\usepackage{tikz}
\usetikzlibrary{arrows.meta, positioning, fit, calc}

\title{Reservoir computing based on multicore fibers}

\author[1,2,3]{Igor Chekhovskoy}
\author[3]{Stanislav Mitsai}
\author[1,2]{Georgiy Patrin}
\author[1,2]{Mikhail Fedoruk}
\author[3]{Alexey Kokhanovskiy}

\affil[1]{Novosibirsk State University, Novosibirsk, 630090, Russia}
\affil[2]{Federal Research Center for Information and Computational Technologies, Novosibirsk, 630090, Russia}
\affil[3]{School of Physics and Engineering, ITMO University, 197101 Saint Petersburg, Russia}

\begin{document}
\maketitle

\begin{abstract}
Photonic reservoir computing offers a hardware-efficient route to processing temporal and sequential data, but delay-based implementations often rely heavily on temporal multiplexing, where long temporal masks are required to generate a sufficiently rich reservoir state. Here we present a numerical proof of concept showing that the spatial degrees of freedom of an active multicore fiber placed inside a delayed optical feedback loop can reduce this dependence on serial temporal encoding. The input signal is encoded by temporal and spatial masks, the pump distribution across the cores controls the reservoir operating point through the core-dependent effective gain and saturation energy, and the detected core intensities serve as readout features for a single trained linear layer. The system is modeled by linearly coupled nonlinear Schr\"odinger equations with saturable gain and solved using a split-step Fourier method. On the Mackey--Glass one-step-ahead prediction benchmark, a seven-core reservoir with equal temporal masks reduces the normalized root mean square error (NRMSE) from 0.5956 for the single-core baseline to 0.0651 at a modulation rate of 40 GHz. At 1 GHz, a seven-core reservoir without temporal masking reaches an NRMSE of 0.0323. These results show that an active multicore fiber can provide both parallel readout channels and a tunable nonlinear transformation, offering a route to photonic reservoirs with reduced reliance on temporal multiplexing.
\end{abstract}

\section{Introduction}

Reservoir computing (RC) has attracted considerable attention as a computational paradigm for processing temporal and sequential data. In contrast to conventional recurrent neural networks, the internal connections of the reservoir are not trained. Instead, the reservoir transforms the input signal into a high-dimensional dynamical state, and only the final linear readout is optimized. This property makes RC especially attractive for physical implementations, because the complex dynamics of an optical, electronic, mechanical, magnetic or other physical system can be used directly as the computational substrate~\cite{Tanaka2019PhysicalReservoirComputing}. The resulting training procedure is simple, while the reservoir itself may provide nonlinear transformation, memory and feature expansion.

Photonics is a particularly promising platform for physical reservoir computing and neuromorphic information processing. Optical systems naturally provide large bandwidth, low latency and several forms of parallelism, including wavelength, polarization, temporal and spatial multiplexing~\cite{Shastri2021PhotonicsAI,McMahon2023PhysicsOpticalComputing}. In wave-based photonic reservoirs, information can be encoded and processed through the propagation of optical fields in a physical medium rather than by explicitly implementing a large number of artificial neurons. In this setting, the useful reservoir transformation is produced by the interplay between linear wave mixing and optical nonlinearity. The main design challenge is therefore not only to obtain a sufficiently high-dimensional optical state, but also to tune the balance between memory, nonlinearity and stability for a given task.

A major line of photonic RC research is based on time-delay reservoirs. In this approach, the input sequence is masked in time and injected into a nonlinear node with delayed feedback. The delayed loop creates a set of virtual nodes, allowing a compact physical system to emulate a high-dimensional recurrent reservoir. Appeltant et al. demonstrated that a single nonlinear node with delayed feedback can solve benchmark tasks including chaotic time-series prediction~\cite{Appeltant2011InformationSystem}. This concept was subsequently implemented in optoelectronic hardware by Paquot et al., who showed real-time information processing with performance comparable to digital methods on practical benchmark tasks~\cite{Paquot2012OptoelectronicComputing}. Later, Duport et al. demonstrated a fully analogue photonic reservoir computer, reducing the need for digital pre- and post-processing and confirming the relevance of analogue photonic dynamics for RC~\cite{Duport2016FullyComputer}. Integrated photonic implementations, including microring resonators with external optical feedback, further illustrate that compact passive or weakly nonlinear optical devices can be used in time-delay reservoir architectures~\cite{Donati:22}. Related SOA-based reservoir-computing schemes have also been proposed, including an upsampling-and-modulation approach using a semiconductor optical amplifier and a photodetector as nonlinear elements~\cite{Manuylovich:24}.
Recent experimental work on semiconductor-laser time-delay reservoirs has further shown that the effective state space can be expanded at the readout level by upsampling and nonlinear transformations of virtual-node responses~\cite{Kovalev2025StateSpaceExpansion}.

Despite these advantages, delay-based RC relies heavily on temporal multiplexing. A large number of virtual nodes is usually obtained by stretching each input symbol over many mask positions. This serial encoding increases the effective duration of one processed symbol and may limit throughput. Recent work has partly addressed this limitation by combining dense temporal encoding with wavelength-division multiplexing in an all-optical reservoir based on a nonlinear amplifying loop mirror~\cite{Aadhi2025ScalablePhotonicRC,DiLauro2026PhotonicsBreakthroughs}. In the present work, we explore a different and complementary route based on spatial degrees of freedom: instead of increasing the reservoir state only by samples along a temporal delay line, one can also use spatial channels, optical modes or fiber cores as parallel components of the state vector.

Time-delay reservoirs create many virtual nodes by dividing each input symbol into a sequence of masked time intervals. Spatial channels offer another way to enlarge the reservoir state because several optical channels can be processed and read out in parallel. However, simply adding parallel channels is not enough. If the channels do not interact, they may produce redundant copies of the same response; if they mix too strongly, their outputs become indistinguishable. A useful spatial reservoir therefore requires controlled excitation, interaction and separate detection of the channels. Multicore fibers are attractive for this purpose because their cores provide discrete spatial channels with controllable coupling and direct access at the output. As the number of cores grows, nonuniform gain, fabrication differences and readout cross-talk must also be controlled.

Recent wave-based optical computing studies have demonstrated the usefulness of spatially extended optical media for machine learning. Large-scale optical reservoir computing has been demonstrated using optical propagation for spatiotemporal chaotic systems prediction~\cite{Rafayelyan2020LargeScaleORC}. Multimode fibers have also been used as scalable optical learning operators, where simultaneous linear and nonlinear interaction of spatial modes forms the computational engine~\cite{Tegin2021ScalableOpticalLearningOperator}. Related works on multimode-fiber-based imaging and computing further support the idea that complex modal propagation can be exploited as a physical transformation for machine learning tasks~\cite{Rahmani2022LearningImageCompute}. More recently, gain-controlled multimode fibers were proposed as tunable nonlinear photonic reservoirs, where the operating regime can be controlled by pump and signal powers, and where the best performance appears in an intermediate regime between an overly linear response and excessive nonlinear randomness~\cite{Marcucci2025GainControlledMMF}. In addition, multimode-fiber laser cavities have recently been considered as nonlinear optical processors, where phase-encoded input patterns undergo high-dimensional transformations through multimode interference and gain-saturation dynamics~\cite{Eslik2026MMFLaserCavities}.

In the present work, we follow this general direction but replace the multimode spatial continuum by a multicore fiber (MCF). MCF technology was originally developed for space-division multiplexed optical communications, where several cores inside the same cladding act as parallel spatial channels~\cite{Richardson2013,SaitohMatsuo2016,Hayashi2011,6093761}. Unlike overlapping modes of a multimode fiber, the MCF cores form discrete channels that can in principle be addressed and detected separately. At the same time, inter-core coupling, Kerr nonlinearity, saturable gain and optical feedback allow these channels to participate in a common recurrent transformation rather than acting only as independent transmission paths. In contrast to a single-core delay reservoir, an MCF reservoir can therefore increase the number of directly accessible readout features without relying only on a long temporal mask.

The main contribution of this work is an active multicore-fiber reservoir in which delayed optical feedback provides memory and the coupled cores provide parallel nonlinear states. During each pass through the MCF, inter-core coupling, Kerr nonlinearity and core-dependent saturable gain transform the input field across the cores. The output intensity of each core is measured separately, while part of the optical field is returned to the next input window. The reservoir state is therefore enlarged by physical spatial channels rather than only by increasing the length of the temporal mask. This architecture differs from the gain-controlled multimode-fiber reservoir of Marcucci et al., where the fiber performs a feed-forward nonlinear transformation and time-series memory is introduced through delay embedding in post-processing~\cite{Marcucci2025GainControlledMMF}. We consider 1-, 7- and 19-core configurations and compare two main encoding strategies: equal temporal masks across cores and spatial-only encoding. This allows us to test whether discrete spatial channels can reduce the required temporal mask length and whether a reservoir based only on spatial input encoding can remain accurate.

\begin{table}[ht]
\centering
\footnotesize
\setlength{\tabcolsep}{3pt}
\renewcommand{\arraystretch}{1.15}
\begin{tabularx}{\textwidth}{>{\RaggedRight\arraybackslash}p{0.19\textwidth} >{\RaggedRight\arraybackslash}p{0.17\textwidth} >{\RaggedRight\arraybackslash}p{0.18\textwidth} >{\RaggedRight\arraybackslash}X}
\hline
Architecture & Memory mechanism & Parallel degrees of freedom & Relation to the present work \\
\hline
Single-node delay reservoir~\cite{Appeltant2011InformationSystem,Paquot2012OptoelectronicComputing} & Physical delayed feedback & Temporal virtual nodes & Compact recurrent architecture; its state dimension is generated mainly by serial temporal masking. \\
Wavelength-parallel delay reservoir~\cite{Aadhi2025ScalablePhotonicRC} & Physical delayed feedback & Wavelength and temporal channels & Reduces the serial bottleneck through spectral rather than spatially resolved fiber channels. \\
Gain-controlled multimode-fiber reservoir~\cite{Marcucci2025GainControlledMMF} & Feed-forward transformation; delay embedding for time-series tasks & Overlapping spatial modes & Provides pump-tunable nonlinear spatial features but no intrinsic physical feedback memory. \\
Active MCF reservoir (this work) & Physical delayed feedback & Discrete cores and optional temporal nodes & Uses core-resolved recurrent states, inter-core mixing, independently controlled gain and direct intensity readout. \\
\hline
\end{tabularx}
\caption{Qualitative positioning of the proposed architecture relative to representative photonic reservoirs. Direct numerical comparison of reported prediction errors is not attempted because the benchmark parameters, prediction horizons, preprocessing and data-splitting protocols differ among studies.}
\label{tab:architecture_comparison}
\end{table}

We evaluate the proposed MCF reservoir on the Mackey--Glass one-step-ahead prediction benchmark. This task tests nonlinear transformation and short-term memory and is widely used for assessing reservoir-computing systems. Within the selected fiber-length range, no spatial-only configuration was physically admissible at 40 GHz, whereas accurate spatial-only regimes were found at 1 GHz. This difference follows from the feedback-loop timing, because one symbol window must be longer than the propagation time through the MCF. The timing constraint is discussed in Sec.~\ref{sec:mg_prediction}. The results show that spatial degrees of freedom in an MCF can partially replace temporal multiplexing for chaotic time-series prediction.

\section{Simulated laser system with delayed feedback loop based on multicore fiber}

\begin{figure}[ht]
    \centering
    \includegraphics[width=0.7\linewidth]{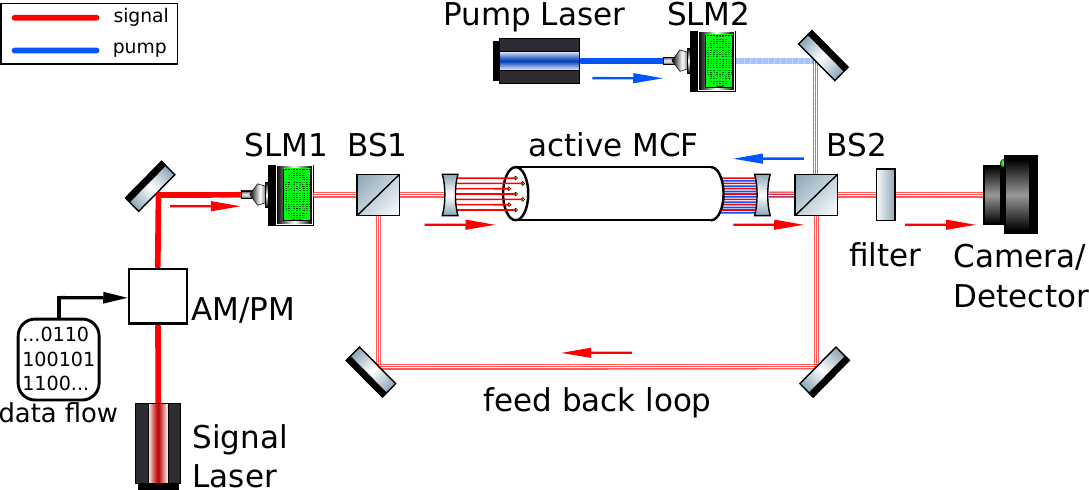}
    \caption{Scheme of the simulated laser system with delayed feedback loop based on multicore fiber.}
    \label{fig:device_scheme}
\end{figure}

Figure~\ref{fig:device_scheme} depicts the principal scheme of an MCF reservoir computer. The proposed reservoir is implemented as an active multicore-fiber system placed inside an optical feedback loop. The general idea follows delay-based photonic reservoir computing, where memory is introduced by feeding a delayed part of the optical output back to the input, while the internal optical dynamics of the system remain fixed and only the final readout is trained~\cite{Appeltant2011InformationSystem,Paquot2012OptoelectronicComputing,Duport2016FullyComputer,Donati:22}. In contrast to conventional single-node delay reservoirs, the present scheme uses the spatial degrees of freedom of an MCF. Therefore, the reservoir state is formed not only by time-multiplexed samples, but also by the optical fields in different coupled cores.

The input signal is modulated in both the temporal and spatial domains before being launched into the MCF. Temporal modulation of the amplitude or phase of a time-series signal may be achieved using standard LiNbO$_3$ modulators. The temporally modulated optical field is then sent to the first spatial light modulator (SLM1), which distributes the signal between the MCF cores. This spatial modulation can be realized using a liquid-crystal light modulator, a deformable-mirror device, or a static diffractive optical element. In the numerical model this operation corresponds to multiplying the scalar input sequence by prescribed temporal and spatial masks. Thus, SLM1 determines which cores are directly excited by the input and how the input symbols are spread over the temporal and spatial degrees of freedom of the reservoir. Similar spatial structuring of optical fields is widely used in spatially multiplexed optical computing and photonic reservoir systems~\cite{Rafayelyan2020LargeScaleORC}.

The pump laser is injected into the active MCF through a separate optical path. To achieve greater tunability, different gain levels can be introduced in individual cores through spatial modulation of the pump laser by means of a second spatial light modulator (SLM2). In the reduced numerical model, the pump field is not propagated explicitly. Its effect is represented by the effective gain coefficients $g_{0,n}$ and the corresponding saturation energies in different cores. This makes it possible to tune the operating point of the active MCF without changing the internal structure of the reservoir. This idea is close to gain-controlled fiber reservoirs, where signal and pump beams are structured in order to control the nonlinear transformation produced by the active medium~\cite{Marcucci2025GainControlledMMF}. In the present work, the pump distribution is treated as an externally controlled parameter and is optimized together with the feedback coefficient, phase shift and fiber length.

Recent experimental work on transverse-mode-division control in an active Yb-doped fiber laser illustrates that spatial degrees of freedom in gain fibers can be engineered by mode-selective components~\cite{Ma2025TransverseModeDivision}. The present proposal requires a different, core-resolved implementation: the signal and pump fields must be mapped onto individual cores, and the output intensities must be registered with sufficiently low cross-talk. The scheme should therefore be regarded as a numerical proof of concept. A practical implementation would additionally require calibration of the launch and detection matrices and stabilization of the feedback phase.

After propagation through the multicore fiber, the signal is divided into two parts. The first part is relaunched through the feedback loop back into the reservoir. The feedback loop introduces memory between consecutive input windows: the output field from the previous pass is delayed, multiplied by the feedback coefficient, and added to the next externally injected signal. The second part is registered by a digital camera and used for digital signal processing. In the simulations, the detected features are obtained from the optical intensities in the cores and then used for training the linear readout.

In this study, we investigate the performance of the MCF reservoir computer in two input-encoding regimes. In the first regime, the same temporal modulation is applied across all cores, while spatial modulation controls the distribution of the input signal between the cores. This configuration is closest to a conventional time-multiplexed delay reservoir, but the signal still propagates through several coupled MCF cores. In the second regime, only spatial modulation is used, and the temporal mask is reduced to a single sample per input symbol. This regime is especially attractive because it can reduce the number of serial temporal-mask positions, but it also provides fewer time-multiplexed virtual nodes and therefore relies more strongly on spatial mixing and feedback memory.

\section{Mathematical model of multicore fiber}

\begin{figure}[ht]
    \centering
    \includegraphics[width=0.7\linewidth]{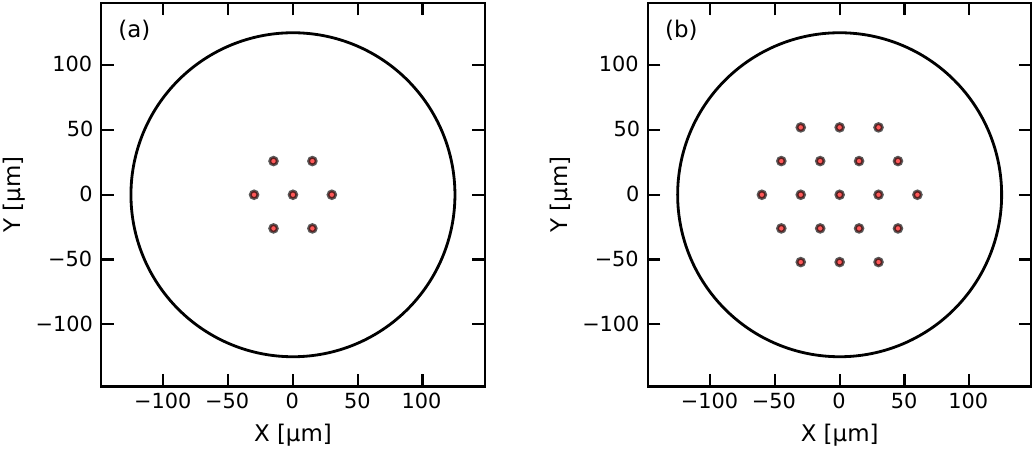}
    \caption{Scheme of considered 7-core (a) and 19-core (b) hexagonal MCFs. The red circles represent the cores and the solid line represents the cladding boundary.}
    \label{fig:MCF_schemes}
\end{figure}

The total electromagnetic field in MCF of length~$L$ with $N$ cores (see Fig.~\ref{fig:MCF_schemes}) can be described as a superposition of optical modes localized at each core~\cite{Mumtaz2013, 6253229, Mumtaz:12}:
\begin{equation}
\label{eq:NLSE}
E(x,y,z,t) = \sum\limits_{n=1}^{N} A_n(z,t)F_n(x-x_n, y-y_n)e^{i(\beta_n z-\omega t)} + cc,
\end{equation}
where $A_n = A_n(z,t)$ are slowly varying pulse envelopes at each individual core with coordinates $(x_n, y_n)$ and propagation constant $\beta_n$, $F_n$ is the spatial mode structure in the $n$-th core, and the ``cc'' denotes an operation of complex conjugation. Also $z \in [0, L]$ is the spatial variable and $t \in [-T/2, T/2]$ is the temporal variable.
The dynamics of envelopes $A_n$ in the presence of saturable gain can be described by a system of linearly coupled NLSE.
The full dimensional model can be written in the form
\begin{equation}
 \label{eq:SysNLSE_full}
 i\frac{\partial A_n}{\partial z}
 =
 -i \beta_1\frac{\partial A_n}{\partial t}
 + \frac{\beta_2}{2}\frac{\partial^2 A_n}{\partial t^2}
 - \gamma |A_n|^2 A_n
 + \frac{i{g_{0,n}}A_n}{2 \left(1 + \frac{E_n(z)}{E_{\text{sat},n}}\right)}
 - \frac{i\alpha_n}{2}A_n
 - \sum_{m=1}^N C_{n,m} A_m
 + A_{\text{ASE},n}.
\end{equation}
Here $\beta_1$ is the inverse group velocity, $\beta_2$ is the group velocity dispersion parameter and $\gamma$ is Kerr nonlinear coefficient, which are constant and equal for different cores for simplicity. The coefficient $g_{0,n} > 0$ describes the effective gain in the $n$-th core, and $\alpha_n$ is the linear loss coefficient. Gain saturation occurs due to the term $1 + \frac{E_n(z)}{E_{\text{sat},n}}$ in the denominator, where
\begin{equation}
\label{eq:core_energy}
E_n(z) = \int\limits_{-T/2}^{T/2} |A_n(z,t)|^2 dt
\end{equation}
represents the optical energy in the $n$-th core at position $z$, and
\begin{equation}
\label{eq:saturation_energy}
E_{\text{sat},n} = P_{\text{sat},n} T
\end{equation}
is the saturation energy corresponding to the computational time window of duration $T$. The matrix $\mathbf C = (C_{n,m})$ defines linear couplings among envelopes $A_n$ in different cores. Component $A_{\text{ASE},n}$ denotes amplified spontaneous emission noise in the active fiber.

The optical and geometrical parameters used in the simulations are calculated from the carrier wavelength and from the MCF geometry. The carrier wavelength is fixed as
$\lambda_0 = 1.55~\mu\mathrm{m}$, at which the dispersion is anomalous, i.e. $\beta_2<0$.
The cores are arranged in a hexagonal geometry, as shown in Fig.~\ref{fig:MCF_schemes}. In the reported 7-core and 19-core configurations, the core radius is $r_c=2.95~\mu\mathrm{m}$, the core pitch is $\Lambda=30~\mu\mathrm{m}$, the numerical aperture is $\mathrm{NA}=0.125$, and the nonlinear refractive index is taken as
$n_2 = 3.2\times 10^{-20}~\mathrm{m^2/W}$.
For these parameters, each individual core operates in the single-mode regime, so only the fundamental LP$_{01}$ spatial mode is retained in the model. The details of the refractive-index model, LP$_{01}$ mode calculation, effective area and coupling-coefficient calculation are given in~\ref{app:mcf_geometry}.

For the representative 7-core geometry, the calculated propagation and nonlinear parameters are
\begin{equation}
\label{eq:computed_beta_values}
\beta_1 = 4892.85~\mathrm{ps/m},
\qquad
\beta_2 = -7.76\times 10^{-3}~\mathrm{ps^2/m},
\qquad
\gamma = 1.718\times 10^{-3}~\mathrm{W^{-1}m^{-1}}.
\end{equation}
The value of $\beta_1$ is used to determine the propagation time through the MCF section and the corresponding free-space delay in the feedback loop. The value of $\beta_2$ is computed as a physical diagnostic parameter and is used when estimating characteristic propagation lengths. However, in the main feedback-loop simulations the dispersive term is switched off, as described below.

The coupling matrix $\mathbf C$ is calculated from the MCF geometry and modal overlap integrals for the actual core positions. The treatment of the finite cladding boundary and the isolated-core LP$_{01}$ approximation is given in~\ref{app:mcf_geometry}. After computing the full matrix, coupling coefficients smaller than $10^{-3}$ of the largest off-diagonal coefficient are neglected. This gives a sparse effective coupling matrix used in the simulations. For the representative 7-core geometry, the nearest-neighbor coupling coefficient is
$C_{n,m} = 5.272~\mathrm{m^{-1}}$,
which corresponds to the coupling length
\begin{equation}
\label{eq:coupling_length_value}
L_c = \frac{\pi}{2C_{n,m}}=0.298~\mathrm{m}.
\end{equation}

The gain and saturation parameters are not treated as fixed material constants. They are determined by the pump power distribution between the cores. In the simulations, the pump power in the $n$-th core, $P_{\mathrm{pump},n}$, is converted into the effective gain coefficient $g_{0,n}$ and the saturation power $P_{\mathrm{sat},n}$ using empirical approximations for an Er-doped active fiber~\cite{Shtyrina2017JOSAB}. Thus, in the reservoir simulations the optimized pump distribution controls both the gain profile and the saturation level of the active MCF. The pump-to-effective-gain and pump-to-saturation relations are described in~\ref{app:pump_gain}.

For each reservoir operating point, the pump powers are converted once into $g_{0,n}$ and $P_{\mathrm{sat},n}$, and these parameters remain fixed during the simulated sequence. Gain saturation is evaluated from the current optical field through $E_n(z)$ in Eqs.~\eqref{eq:SysNLSE_full} and~\eqref{eq:SysNLSE}. During the feedback simulation, the propagation equation is solved successively for round trips indexed by $r$. This index is omitted in the propagation equations for compactness: at round trip $r$, $A_n(z,t)$ and $E_n(z)$ denote $A_n^{(r)}(z,t)$ and $E_n^{(r)}(z)$, respectively. Since the feedback field changes the input to the next pass, $E_n(z)$ and therefore the saturation term vary both along the MCF and between successive round trips. The model assumes a quasi-stationary gain response and does not resolve population-inversion dynamics, pump depletion, the upper-state lifetime or relaxation oscillations.

In the numerical model used below, several simplifications are introduced. First, the propagation equation is written in the retarded time frame moving with the group velocity; therefore, the term with $\beta_1$ is removed from the propagation equation, while the corresponding group delay is still used when propagation times are evaluated.
Second, distributed passive propagation losses inside the short MCF section are neglected, so $\alpha_n = 0$ in the propagation equation used below. This approximation is justified because the considered MCF length is of the order of meters, whereas passive silica-fiber attenuation at the carrier wavelength corresponds to a loss length of many kilometers. Losses accumulated outside the active MCF section, including beam splitters, coupling losses, free-space propagation and reinjection efficiency of the feedback loop, are included in the lumped feedback coefficient $\kappa$. The coefficient $g_{0,n}$ is therefore treated as an effective gain coefficient controlled by the pump power rather than as a separately calibrated microscopic material parameter.
Third, the main numerical results are obtained in the noise-free regime. Therefore, $A_{\text{ASE},n}=0$ in the propagation equation. When ASE noise is explicitly enabled, it is not treated as a distributed source term. Instead, at each feedback round trip, it is added at the end of the MCF section as lumped additive complex Gaussian noise. The optional ASE model and the effective output OSNR estimate are given in~\ref{app:ase_osnr}.

The dispersion term is retained in the general model, but in the reservoir simulations with feedback loop we set $\beta_2 = 0$.
This approximation was used because the delayed feedback loop requires repeated propagation of mutually dependent fields from previous round trips, and inclusion of dispersion significantly increases the computational cost of the iterative loop simulation. At the same time, for the considered operating regimes the dispersion length is much larger than the active MCF length. For example, in the representative 7-core regime the estimated dispersion length is $L_D=3.18\times 10^5~\mathrm{m}$, whereas the gain and coupling lengths are below one meter. The corresponding characteristic lengths and the longitudinal step selection are given in~\ref{app:numerical_propagation}.

With these assumptions, the propagation model used in the main simulations is
\begin{equation}
 \label{eq:SysNLSE}
 i\frac{\partial A_n}{\partial z}
 =
 - \gamma |A_n|^2 A_n
 + \frac{i{g_{0,n}}A_n}{2 \left(1 + \frac{E_n(z)}{E_{\text{sat},n}}\right)}
 - \sum_{m=1}^N C_{n,m} A_m,
 \qquad n=1,\dots,N.
\end{equation}

\section{Numerical algorithm for modeling multicore fiber reservoir with feedback loop}

To solve the system~\eqref{eq:SysNLSE}, we applied the numerical method from~\cite{Patrin2024ModificationGain}.
The algorithm is a variant of the split-step Fourier method (SSFM)~\cite{TahaAblowitz1984, Agrawal, ChekhovskoyEtAl_JoCP_2017}. In the general implementation, the propagation step can include the dispersion term with $\beta_2$. However, in all feedback-loop reservoir simulations reported below we use $\beta_2=0$. In this case the linear step does not require a dispersive Fourier multiplier and only the inter-core coupling part remains in the linear operator. The longitudinal step (by $z$) is selected using characteristic propagation lengths. The detailed propagation scheme, including the Strang decomposition, analytical saturable-gain nonlinear step and longitudinal step selection, is given in~\ref{app:numerical_propagation}.

The feedback loop is implemented by adding a delayed copy of the output field from the previous pass through the MCF to the new external input signal. In the discrete model this can be written as
\begin{equation}
\label{eq:feedback_loop}
A_{n,\text{in}}^{(r+1)}(t)
=
A_{n,\text{ext}}^{(r+1)}(t)
+
\kappa \exp(i\Delta\phi_f)
A_{n,\text{out}}^{(r)}(t-T_f),
\end{equation}
where $A_{n,\text{in}}^{(r+1)}$ is the field launched into the $n$-th core at the next pass, $A_{n,\text{ext}}^{(r+1)}$ is the new externally injected signal, and $A_{n,\text{out}}^{(r)}$ is the output field after the previous pass through the MCF. The index $r$ enumerates consecutive feedback windows, $\kappa$ is the feedback amplitude coefficient, $\Delta\phi_f$ is the effective feedback phase shift, and $T_f$ is the feedback delay represented by one computational window. The relation between the computational window, MCF group delay and external free-space delay is described in~\ref{app:feedback_timing}.

In the numerical implementation, the output field from the previous window is stored, multiplied by $\kappa \exp(i\Delta\phi_f)$, and added to the next input window. The absolute optical carrier phase accumulated along the fiber and free-space parts of the loop is not calculated explicitly, because in an experimental system this phase is sensitive to the optical path length and environmental perturbations. Therefore, only the feedback amplitude and an effective phase shift $\Delta\phi_f$ are used as controllable parameters.

\section{Reservoir computing pipeline}

The reservoir computing pipeline converts the optical output of the MCF system into a supervised learning model. For each input window, the simulated optical system produces a set of output fields in the MCF cores. The reservoir state vector $\mathbf{x}_t \in \mathbb{R}^{D}$ is formed from the detected optical intensities at symbol index $t$. In the considered configurations, these features include the intensity samples collected from different cores and, when temporal masking is used, from different mask positions inside the same input symbol. For a fixed operating regime, the optical transformation performed by the MCF system is kept unchanged during training; only the final linear readout is optimized.

The input encoding stage is described by two mask factors. The scalar input symbol $u_t$ is multiplied by a temporal factor $m_j$, where $j=1,\dots,M$, and by a spatial factor $s_n$, where $n=1,\dots,N$. In the numerical model, this gives the encoded optical input in the form
\begin{equation}
\label{eq:input_encoding}
A_{n,j}^{\mathrm{in}}
\propto
s_n m_j u_t.
\end{equation}
The temporal factor determines how one input symbol is distributed over the mask positions inside the symbol window, while the spatial factor determines how the same symbol is distributed between different MCF cores. Setting $M=1$ disables temporal masking, and choosing uniform $s_n$ disables spatial modulation of the input distribution. Thus, the same formulation covers temporal and spatial masking, spatial-only encoding, and the limiting temporal-only case.

\begin{figure}[ht]
    \centering
    \resizebox{\linewidth}{!}{%
        \begin{tikzpicture}[
    font=\scriptsize,
    >=Latex,
    block/.style={
        draw,
        rounded corners=2pt,
        align=center,
        minimum height=0.72cm,
        minimum width=1.78cm,
        inner sep=3pt
    },
    smallblock/.style={
        draw,
        rounded corners=2pt,
        align=center,
        minimum height=0.55cm,
        minimum width=1.65cm,
        inner sep=2pt
    },
    wideblock/.style={
        draw,
        rounded corners=2pt,
        align=center,
        minimum height=0.76cm,
        minimum width=2.18cm,
        inner sep=3pt
    },
    arrow/.style={->, thick},
    note_arrow/.style={->, thick, dashed},
    dashedbox/.style={draw, dashed, rounded corners=2pt, inner sep=4pt}
]

% -------------------------
% Optical encoding and propagation
% -------------------------

\node[block] (input) {input symbol\\$u_t=x_t$};

\node[wideblock, right=0.62cm of input] (encoder) {
    input encoding\\[1pt]
    $A_{n,j}^{\mathrm{in}}\propto s_n m_j u_t$
};

\node[smallblock, above=0.42cm of encoder] (tm) {
    temporal factor\\
    $m_j,\ j=1,\dots,M$\\
    $M=1$ disables it
};

\node[smallblock, below=0.42cm of encoder] (sm) {
    spatial factor\\
    $s_n,\ n=1,\dots,N$\\
    uniform $s_n$ disables it
};

\node[block, right=0.62cm of encoder] (masked) {
    encoded optical\\
    input $A_{n,j}^{\mathrm{in}}$
};

\node[block, right=0.62cm of masked] (mcf) {
    active MCF\\
    fixed optical\\
    transformation
};

\node[block, right=0.62cm of mcf] (detector) {
    detected\\
    intensities\\
    $I_{n,j}(t)=|A_{n,j}^{\mathrm{out}}|^2$
};

\node[smallblock, above=0.60cm of mcf] (feedback) {
    delayed feedback\\
    $\kappa e^{i\Delta\phi_f}$
};

\node[smallblock, below=0.60cm of mcf] (pump) {
    pump profile\\
    $g_{0,n},\ E_{\mathrm{sat},n}$
};

\draw[arrow] (input) -- (encoder);
\draw[arrow] (tm) -- (encoder);
\draw[arrow] (sm) -- (encoder);
\draw[arrow] (encoder) -- (masked);
\draw[arrow] (masked) -- (mcf);
\draw[arrow] (mcf) -- (detector);

\draw[arrow] (mcf.north) -- (feedback.south);
\draw[arrow] (feedback.west) -| ($(masked.north)+(0.12,0)$);
\draw[arrow] (pump.north) -- (mcf.south);

\node[dashedbox, fit=(tm) (sm) (encoder)] (encodingbox) {};
\node[above=0.04cm of encodingbox] {mask configuration};

\node[wideblock, below=0.62cm of encodingbox] (modes) {
    regimes studied\\[1pt]
    equal temporal masks across cores: $M>1$\\
    spatial-only encoding: $M=1$, nonuniform $s_n$
};

\draw[note_arrow] (modes.north) -- (encodingbox.south);

% -------------------------
% State construction and readout
% -------------------------

\node[wideblock, below=1.12cm of detector] (state) {
    state vector\\[1pt]
    $\mathbf{x}_t=\operatorname{vec}\{I_{n,j}(t)\}$\\
    $\mathbf{x}_t\in\mathbb{R}^{NM}$
};

\node[wideblock, right=0.92cm of state] (readout) {
    trained linear\\
    readout\\[1pt]
    $\widehat{y}_t=
    \mathbf{w}_{\mathrm{out}}^\top
    \mathbf{x}_t+b_{\mathrm{out}}$
};

\node[wideblock, above=0.52cm of readout] (target) {
    Mackey--Glass\\
    prediction target\\[1pt]
    $y_t=x_{t+1}$
};

\draw[arrow] (detector) -- (state);
\draw[arrow] (state) -- (readout);
\draw[arrow] (target.south) -- (readout.north);

\end{tikzpicture}
    }
    \caption{Reservoir-computing pipeline used in the MCF system. The scalar input symbol is encoded by temporal and spatial masks, propagated through the fixed active-MCF reservoir with delayed feedback, and converted into a state vector from detected core intensities. Only the final linear readout is trained.}
    \label{fig:rc_pipeline}
\end{figure}
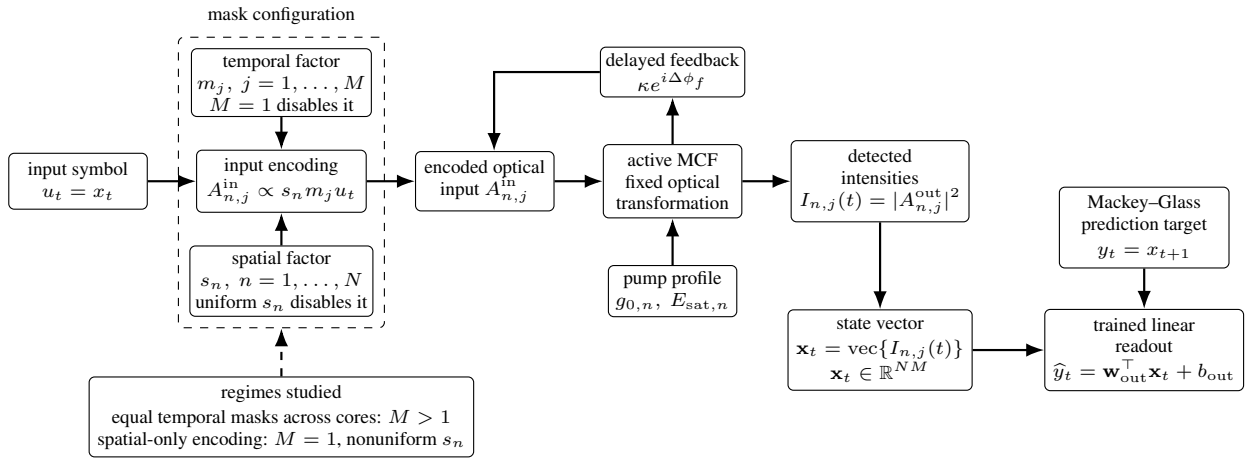

In the reported simulations below, we focus on two main input-encoding regimes. In the temporal-masking regime, the same temporal mask of length $M$ is applied to all cores, while the input can still be distributed between the cores by spatial weights. In the spatial-only regime, the input is distributed between the cores by spatial weights alone. In all reported simulations, the random masks are generated from a uniform distribution and centered before scaling. The detected reservoir features are optical intensities only; complex optical fields are not used directly in the readout. The complete processing chain, from input encoding to the final trained readout, is summarized in Fig.~\ref{fig:rc_pipeline}.

For intensity readout, the feature dimension is
\begin{equation}
\label{eq:feature_dimension}
D = N M,
\end{equation}
where $N$ is the number of cores and $M$ is the number of temporal mask positions.

We consider the Mackey--Glass chaotic time-series prediction benchmark. The Mackey--Glass system is a classical delayed nonlinear system introduced to model oscillatory and chaotic dynamics in physiological control systems~\cite{MackeyGlass1977}. This benchmark probes nonlinear transformation and short-term memory and is commonly used in reservoir-computing studies~\cite{Appeltant2011InformationSystem,Lukosevicius2012PracticalGuide,Aadhi2025ScalablePhotonicRC}. It is therefore suitable for assessing whether the MCF reservoir can replace part of the temporal masking by spatial degrees of freedom.

The Mackey--Glass sequence is generated from the delay differential equation
\begin{equation}
\label{eq:mackey_glass}
\frac{dx}{dt}
=
\frac{\beta x(t-\tau)}
{1+x^n(t-\tau)}
-
\gamma x(t),
\end{equation}
where we use the standard chaotic-regime parameters
$\tau=17$, $n=10$, $\beta=0.2$, $\gamma=0.1$.
The continuous trajectory is sampled with time step $\Delta t=1$, and the one-step-ahead prediction task is defined as
\begin{equation}
\label{eq:mg_prediction_task}
u_t = x_t,
\qquad
y_t = x_{t+1}.
\end{equation}

We first generate an initial warm-up part of length $N_{\mathrm{warm}} = 500$
and discard it before training or evaluation. This initial segment is used only to remove the dependence on arbitrary initial conditions of the benchmark sequence and of the reservoir response. After this warm-up interval, the next
$N_{\mathrm{seq}} = 10000$ symbols are retained for reservoir-computing evaluation. The scalar input and target sequences are normalized before optical encoding. The retained symbols are then processed in chronological order and split into three non-overlapping subsets in the ratio $8:1:1$:
$N_{\mathrm{train}}=8000$, $N_{\mathrm{val}}=1000$, $N_{\mathrm{test}}=1000$. The corresponding index sets are
\begin{equation}
\label{eq:chronological_split}
\mathcal{T}_{\mathrm{train}}
=
\{1,\dots,N_{\mathrm{train}}\},
\qquad
\mathcal{T}_{\mathrm{val}}
=
\{N_{\mathrm{train}}+1,\dots,N_{\mathrm{train}}+N_{\mathrm{val}}\},
\end{equation}
\begin{equation}
\label{eq:chronological_split_test}
\mathcal{T}_{\mathrm{test}}
=
\{N_{\mathrm{train}}+N_{\mathrm{val}}+1,\dots,N_{\mathrm{seq}}\}.
\end{equation}
The training subset was used to fit the linear readout for each fixed reservoir configuration. The validation subset was used to select the ridge regularization parameter and to compare different reservoir operating regimes. The test subset was not accessed during model development and was used only once, after all hyperparameters had been fixed, in order to obtain an unbiased estimate of the final predictive performance.

\begin{figure}[ht!]
    \centering
    \includegraphics[width=\linewidth]{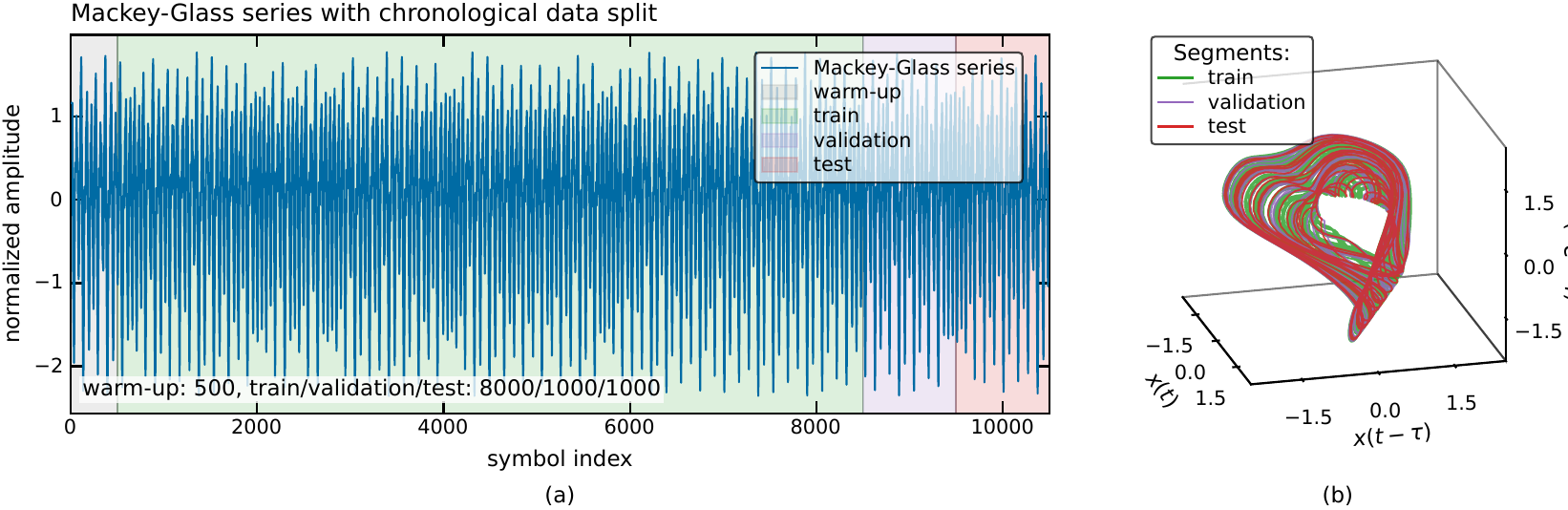}
    \caption{Mackey--Glass benchmark data. (a) Normalized Mackey--Glass time series with warm-up, training, validation and test regions used in the chronological split. (b) Mackey--Glass attractor projection in delay coordinates
$\left(x(t), x(t-\tau), x(t-2\tau)\right)$.}
    \label{fig:mg_overview}
\end{figure}

Figure~\ref{fig:mg_overview} illustrates the Mackey--Glass sequence used in the benchmark and the chronological split into warmup, training, validation and test regions. The attractor projection is shown to indicate the chaotic regime of the generated series.

Such a chronological splitting strategy is essential for time-series prediction problems. In contrast to random reshuffling, it preserves the temporal ordering of the data and prevents information leakage from future observations into the past. For dependent data, ignoring the temporal structure during model selection may lead to overly optimistic error estimates. Therefore, the validation sequence was always split into contiguous blocks when selecting the regularization parameter, and the final test subset was kept separate from all hyperparameter selection steps. This is consistent with the standard reservoir-computing setting, where the recurrent or delayed reservoir dynamics are fixed and only the output layer is trained~\cite{Lukosevicius2012PracticalGuide}.

The readout features were standardized using the training subset only. More precisely, after chronological splitting into training, validation, and test subsets, the feature-wise mean and standard deviation were computed from the training subset, and the same statistics were then applied to the validation and test subsets. Features with nearly zero variance were left unchanged by setting the corresponding standard deviation to unity. This prevents numerical instability without introducing information from validation or test data into the feature normalization stage.

Let $y_t$ be the target value corresponding to the reservoir state $\mathbf{x}_t$. The linear readout was obtained by ridge regression with an intercept term,
\begin{equation}
\label{eq:ridge_readout}
(\mathbf{w}_{\mathrm{out}},b_{\mathrm{out}})
=
\arg\min_{\mathbf{w},b}
\sum_{t \in \mathcal{T}_{\mathrm{train}}}
\left(
y_t - \mathbf{w}^{\top}\mathbf{x}_t - b
\right)^2
+
\alpha \|\mathbf{w}\|_2^2,
\end{equation}
where $\alpha > 0$ is the regularization parameter. After training, the predicted value is denoted by $\widehat{y}_t$ and is computed as
\begin{equation}
\label{eq:readout_prediction}
\widehat{y}_t
=
\mathbf{w}_{\mathrm{out}}^{\top}\mathbf{x}_t
+
b_{\mathrm{out}}.
\end{equation}
Only the readout weights were trained, while the reservoir dynamics and masking procedure remained fixed. Ridge regularization is used to stabilize the readout in the presence of high-dimensional and correlated reservoir features~\cite{Hastie2009Elements,Hastie2020Ridge}.

The prediction error is reported as normalized root mean square error,
\begin{equation}
\label{eq:nrmse}
\mathrm{NRMSE}
=
\sqrt{
\frac{
\sum_t (y_t-\widehat{y}_t)^2
}{
\sum_t (y_t-\overline{y})^2
}
},
\end{equation}
where $\overline{y}$ is the mean of the target values on the corresponding subset.

The regularization parameter $\alpha$ was selected using the one-standard-error rule~\cite{Hastie2009Elements,Friedman2010Glmnet}. For each candidate value $\alpha_j$ from a logarithmic grid, the readout was fitted on the training subset and evaluated on $K=8$ consecutive validation blocks. Let
\begin{equation}
\label{eq:alpha_block_errors}
e_k(\alpha_j)
=
\mathrm{NRMSE}_{\mathcal{V}_k}(\alpha_j),
\qquad
k=1,\dots,K,
\end{equation}
where $\mathcal{V}_k$ is the $k$-th validation block and the normalization factor is computed from the full validation subset. The mean validation error and its standard error were computed as
\begin{equation}
\label{eq:alpha_mean_se}
\overline{e}(\alpha_j)
=
\frac{1}{K}
\sum_{k=1}^{K} e_k(\alpha_j),
\qquad
\mathrm{SE}(\alpha_j)
=
\sqrt{
\frac{1}{K(K-1)}
\sum_{k=1}^{K}
\left(
e_k(\alpha_j)-\overline{e}(\alpha_j)
\right)^2
}.
\end{equation}
Then the best validation value was found as
\begin{equation}
\label{eq:alpha_min_index}
j_{\min}
=
\arg\min_j \overline{e}(\alpha_j),
\end{equation}
and the selected regularization parameter was chosen as
\begin{equation}
\label{eq:alpha_1se_rule}
\alpha_{\mathrm{1SE}}
=
\max
\left\{
\alpha_j:
\overline{e}(\alpha_j)
\le
\overline{e}(\alpha_{j_{\min}})
+
\mathrm{SE}(\alpha_{j_{\min}})
\right\}.
\end{equation}
Thus, among all readouts whose mean validation error is statistically close to the best observed value, we choose the most regularized one. This criterion intentionally shifts the solution toward a smoother and more stable readout, reduces sensitivity to small fluctuations of the validation curve, and suppresses unnecessarily large readout weights. The grid for $\alpha$ was chosen automatically from the singular values of the training design matrix and was restricted to avoid both nearly unregularized ill-conditioned solutions and excessively regularized degenerate readouts.

After selecting $\alpha_{\mathrm{1SE}}$, the readout was refitted on the full training subset using the selected regularization parameter. The validation score was then used to compare different reservoir operating regimes, such as fiber length, feedback strength, phase shift, mask type, input scaling and pump distribution. Only after fixing all such choices was the final model evaluated on the test subset.

The numerical implementation used to generate the reservoir-computing results is available as part of the \texttt{fiberprop} repository, in the \texttt{scripts/mcf\_reservoir\_computing} directory~\cite{ChekhovskoyFiberpropMCFRC}. The calculations were carried out in Python.

\section{Results}

The results are organized around the Mackey--Glass one-step-ahead prediction benchmark. The main goal is to determine whether spatial degrees of freedom in MCF can reduce the required temporal mask length compared with a single-core delay reservoir. We first compare operating regimes with equal temporal masks across cores at 40 GHz and then consider spatial-only encoding at 1 GHz. In all reported benchmark runs, the generated sequence length was fixed to 10000 symbols. Direct numerical ranking against published reservoir-computing errors is not attempted, because the literature uses different Mackey--Glass parameters, prediction horizons, preprocessing, feature dimensions and data-splitting protocols. The validation and test NRMSE values reported below are therefore used for controlled comparisons within the common protocol of the present study.

\subsection{Operating regimes and core-count comparison}

The operating point of the reservoir is controlled by the feedback amplitude coefficient $\kappa$, the effective feedback phase shift $\Delta \phi_f$, the fiber length $L$, the temporal mask size $M$, the input scaling factor $s_{\mathrm{in}}$ and the pump distribution across the cores. The pump distribution determines the gain profile inside the active MCF and therefore affects the balance between amplification, gain saturation, Kerr nonlinearity and inter-core mixing. In the present simulations these parameters were selected by numerical search using the validation subset, while the readout regularization was selected by the one-standard-error procedure described in the previous section.

For each fixed fiber geometry and input-encoding scheme, we optimized the external reservoir-control parameters rather than the internal optical equations. The searched parameters were
\begin{equation}
\label{eq:searched_parameters}
\theta
=
\left(
\kappa,
\Delta \phi_f,
L,
s_{\mathrm{in}},
M,
d_{\mathrm{fb}},
P_{\mathrm{pump},1},
\dots,
P_{\mathrm{pump},N}
\right),
\end{equation}
where $s_{\mathrm{in}}$ is the input scaling factor, $M$ is the temporal mask size, $d_{\mathrm{fb}}$ is the feedback-delay factor measured in input symbols, and $P_{\mathrm{pump},n}$ is the pump power assigned to the $n$-th core. The pump powers were then converted into the effective gain coefficients $g_{0,n}$ and saturation powers $P_{\mathrm{sat},n}$ using the empirical pump-to-effective-gain relations described in the mathematical model section. Thus, the optimization was performed in experimentally meaningful control variables: feedback strength, feedback phase, fiber length, input scaling, mask size, delay and pump powers.

\begin{table}[ht]
\centering
\begin{tabular}{l c l}
\hline
Parameter & Symbol & Search range \\
\hline
Feedback amplitude coefficient & $\kappa$ & $[0.5,\,0.99]$ \\
Feedback phase shift & $\Delta \phi_f$ & $[0,\,2\pi]$ \\
Fiber length & $L$ & $[0.01,\,1.5]~\mathrm{m}$ \\
Input scaling factor & $s_{\mathrm{in}}$ & $[10^{-4},\,1]$ \\
Temporal mask size & $M$ & $\{5,\dots,1000\}$ \\
Feedback-delay factor & $d_{\mathrm{fb}}$ & $\{1,\dots,20\}$ \\
Pump power in each core & $P_{\mathrm{pump},n}$ & $[0,\,3]~\mathrm{W}$ \\
\hline
\end{tabular}
\caption{Search ranges for the reservoir operating parameters. Pump powers below the positive-saturation threshold in Eq.~\eqref{eq:psat_from_pump} were treated as inactive pump channels. For the spatial-only encoding regime, the temporal mask size was fixed to $M=1$.}
\label{tab:search_space}
\end{table}

The numerical search ranges used for the main reported simulations are summarized in Table~\ref{tab:search_space}. The same ranges were used for the 1-core, 7-core and 19-core comparisons, with the number of pump-power variables adapted to the number of cores. For the spatial-only searches discussed below, the mask size and feedback-delay factor were fixed to $M=1$ and $d_{\mathrm{fb}}=1$, respectively.

The lower active pump-power bound corresponds to the onset of positive saturation power in the empirical pump-to-saturation fit, while the upper bound corresponds to the assumed available pump-power range per core. Values below this threshold were used only to allow an optimized solution to effectively switch off pumping in selected cores. This parametrization makes the gain and saturation levels dependent on the same physical control variable, instead of treating $g_{0,n}$ and $P_{\mathrm{sat},n}$ as unrelated fitting parameters. In this sense, the optimized pump distribution represents the action of SLM2 in the simulated system~(see Fig.~\ref{fig:device_scheme}).

For each candidate operating point, the reservoir was simulated on the benchmark input sequence, the readout features were extracted from the detected core intensities, and the linear readout was trained using the training subset. The ridge parameter was selected using the validation-based one-standard-error rule described above. The resulting validation NRMSE was then used as the objective value for comparing different operating points. The test subset was not used during this optimization. After fixing the best operating regimes, the final performance was evaluated on the held-out test subset.

The search over operating regimes was performed using Optuna~\cite{Akiba2019Optuna}. Each trial included a complete reservoir simulation, construction of the readout features and evaluation of the validation NRMSE. Trials with a non-positive external-loop delay, a failed simulation or a non-finite objective value were discarded. The tree-structured Parzen estimator sampler was used with the constant-liar strategy for parallel trials, multivariate and grouped sampling enabled, and the mask seed used as the sampler seed. The chronological training, validation and test boundaries were fixed throughout the search, as were the eight validation blocks used to select the ridge parameter. The optimization was performed in repeated time-limited runs on a computing cluster. The searches were stopped manually when improvements in the best validation NRMSE became infrequent relative to the computational cost of further trials. No common fixed number of trials was imposed. The reported regimes are the best configurations found within the ranges of Table~\ref{tab:search_space}, rather than proven global optima, and the test subset was evaluated only after the operating regime had been fixed. The mask realization was fixed within each search. Robustness to independent mask realizations and parameter perturbations was not quantified.

A critical parameter to control is the phase of the feedback signal ($\Delta \phi_f$)~\cite{Donati:22}. This can be experimentally realized by precisely adjusting the laser wavelength or using a piezoelectric phase shifter in the feedback loop. In fiber-optic feedback delay lines, it is necessary to compensate for temperature fluctuations and mechanical vibrations, which could otherwise lead to a phase shift over time. In the numerical model we therefore treat $\Delta \phi_f$ as an effective controllable parameter rather than calculating the absolute carrier phase accumulated over the full optical path.

Following the logic of gain-controlled fiber reservoirs, we interpret the best operating regions as a balance between insufficient nonlinearity and excessive saturation. In the low-gain or weak-input regime, the reservoir response is close to linear and does not generate sufficiently rich nonlinear features. In the strongly saturated regime, the response becomes less sensitive to the input and may reduce the useful dimensionality of the readout matrix. Therefore, the useful operating point is expected to lie in an intermediate regime where gain, saturation, Kerr nonlinearity and inter-core coupling jointly produce separable but stable features. This interpretation is consistent with recent analysis of gain-controlled multimode-fiber reservoirs, where the best performance was associated with intermediate nonlinear regimes and with the properties of the readout matrix~\cite{Marcucci2025GainControlledMMF}.

\subsection{Mackey--Glass prediction}
\label{sec:mg_prediction}

The Mackey--Glass chaotic time series was generated according to Eqs.~\eqref{eq:mackey_glass}--\eqref{eq:mg_prediction_task}. The task considered here is one-step-ahead prediction. This benchmark requires both nonlinear transformation and memory, and therefore provides a direct test of the proposed MCF-based reservoir. Spatial-only encoding was examined at both 40 and 1 GHz. Within the selected fiber-length range, no spatial-only configuration was physically admissible at 40 GHz, whereas admissible and accurate regimes were found at 1 GHz. The quasi-stationary gain model does not include the upper-state lifetime or relaxation oscillations, so the difference in the present simulations is governed by the feedback-loop timing rather than by these effects. For the spatial-only searches reported here, $M=1$ and $d_{\mathrm{fb}}=1$, so $T_{\mathrm{win}}=1/f_{\mathrm{mod}}$. The condition $T_{\mathrm{free}}=T_{\mathrm{win}}-\beta_1L>0$ then limits the MCF length to $L<5.1~\mathrm{mm}$ at 40 GHz and $L<0.204~\mathrm{m}$ at 1 GHz. Since the fiber-length search started at $0.01~\mathrm{m}$, the 40 GHz spatial-only configuration was not physically admissible within the selected range. At 1 GHz, the best regime used $L=0.120~\mathrm{m}$ and left a positive external-loop delay. Reducing the modulation rate therefore made it possible to use a sufficiently long MCF section for inter-core coupling and gain to form useful spatial states. Other rate-dependent effects would require a separate sweep with matched loop timing.

Table~\ref{tab:temporal_masks_40ghz_best} summarizes the best operating regimes obtained at 40 GHz for reservoirs with equal temporal masks across cores. The 1-core configuration is used as a baseline corresponding to a single active fiber channel with delayed feedback. The 7-core and 19-core configurations test the effect of additional spatial degrees of freedom introduced by the multicore geometry.

\begin{table}[ht]
\centering
\resizebox{\textwidth}{!}{%
\begin{tabular}{c c c c c c c c c c c}
\hline
$N_{\mathrm{cores}}$ & $\kappa$ & $L$, m & $P_{\mathrm{pump}}$, W &
$s_{\mathrm{in}}$ & $\Delta \phi_f$ & $M$ &
$d_{\mathrm{fb}}$ & $\alpha$ & validation NRMSE & test NRMSE \\
\hline
1 &
0.8529 &
1.214 &
\parbox[t]{2.2cm}{1.594} &
0.9491 &
6.2200 &
787 &
1 &
483.265 &
0.596 &
0.625 \\
\hline
7 &
0.6900 &
0.281 &
\parbox[t]{4.8cm}{0.487, 1.257, 0.247, 0.987\\
0.0168, 0.277, 0.537} &
0.0931 &
4.8112 &
102 &
1 &
6675 &
0.0651 &
0.0723 \\
\hline
19 &
0.5870 &
0.281 &
\parbox[t]{6cm}{0.066, 1.626, 1.826, 0.096, 2.116, 2.606\\
2.146, 2.276, 0.136, 2.156, 1.336, 1.506\\
1.966, 0.376, 2.446, 0.286, 0.196, 1.236\\
1.086} &
0.1931 &
1.4820 &
102 &
1 &
5469.86 &
0.0611 &
0.0593 \\
\hline
\end{tabular}%
}
\caption{Best Mackey--Glass regimes at 40 GHz for reservoirs with equal temporal masks across cores. Validation NRMSE was used for model selection, whereas test NRMSE was evaluated only after the operating regime had been fixed.}
\label{tab:temporal_masks_40ghz_best}
\end{table}

The comparison shows that increasing the number of cores improves the prediction accuracy in the considered search space. The improvement from 1 to 7 cores is substantial, indicating that spatial coupling and gain redistribution provide additional useful reservoir features. The transition from 7 to 19 cores gives an additional improvement in the best found 40 GHz regime. This supports the main idea that spatial degrees of freedom in MCF can reduce the burden on temporal masking.

At 1 GHz, accurate spatial-only regimes were found. Their reservoir states are formed by the intensities in different cores and by the feedback dynamics. The best found regimes are summarized in Table~\ref{tab:spatial_only_1ghz_best}.

\begin{table}[ht]
\centering
\resizebox{\textwidth}{!}{%
\begin{tabular}{c c c c c c c c c c c}
\hline
$N_{\mathrm{cores}}$ & $\kappa$ & $L$, m & $P_{\mathrm{pump}}$, W &
$s_{\mathrm{in}}$ & $\Delta \phi_f$ & $M$ &
$d_{\mathrm{fb}}$ & $\alpha$ & validation NRMSE & test NRMSE \\
\hline
7 &
0.8961 &
0.120 &
\parbox[t]{4.8cm}{2.05, 1.63, 1.38, 0.136\\
1.39, 2.31, 2.47} &
0.2700 &
1.2580 &
1 &
1 &
18.084 &
0.0323 &
0.0326 \\
\hline
19 &
0.8810 &
0.120 &
\parbox[t]{6cm}{0.836, 2.446, 1.426, 2.166, 0.446, 1.316\\
2.606, 0.906, 1.956, 1.496, 2.066, 2.216\\
0.426, 0.136, 1.206, 1.646, 2.186, 1.856\\
0.786} &
0.4401 &
4.1610 &
1 &
1 &
1.08 &
0.0157 &
0.0147 \\
\hline
\end{tabular}%
}
\caption{Best Mackey--Glass regimes at 1 GHz for spatial-only encoding. Validation NRMSE was used for model selection, whereas test NRMSE was evaluated only after the operating regime had been fixed.}
\label{tab:spatial_only_1ghz_best}
\end{table}

The best 7-core and 19-core regimes use $L=0.120~\mathrm{m}$, which is compatible with the positive-delay condition discussed above. These results demonstrate that the coupled cores and feedback loop can provide useful reservoir states without temporal masking when the loop timing permits a sufficiently long active-fiber section.

\subsection{Mackey--Glass: 1-core baseline}

As a baseline, we first considered a single-core active fiber with delayed feedback. This configuration preserves temporal masking, the feedback mechanism and the saturable-gain nonlinearity, but does not include inter-core coupling or parallel spatial readout channels. The reservoir state is therefore formed solely from temporal intensity samples of a single optical channel. This provides a direct reference for assessing whether the spatial degrees of freedom of an MCF improve the prediction of the Mackey--Glass dynamics.

\begin{figure}[!htbp]
    \centering
    \includegraphics[width=1\linewidth]{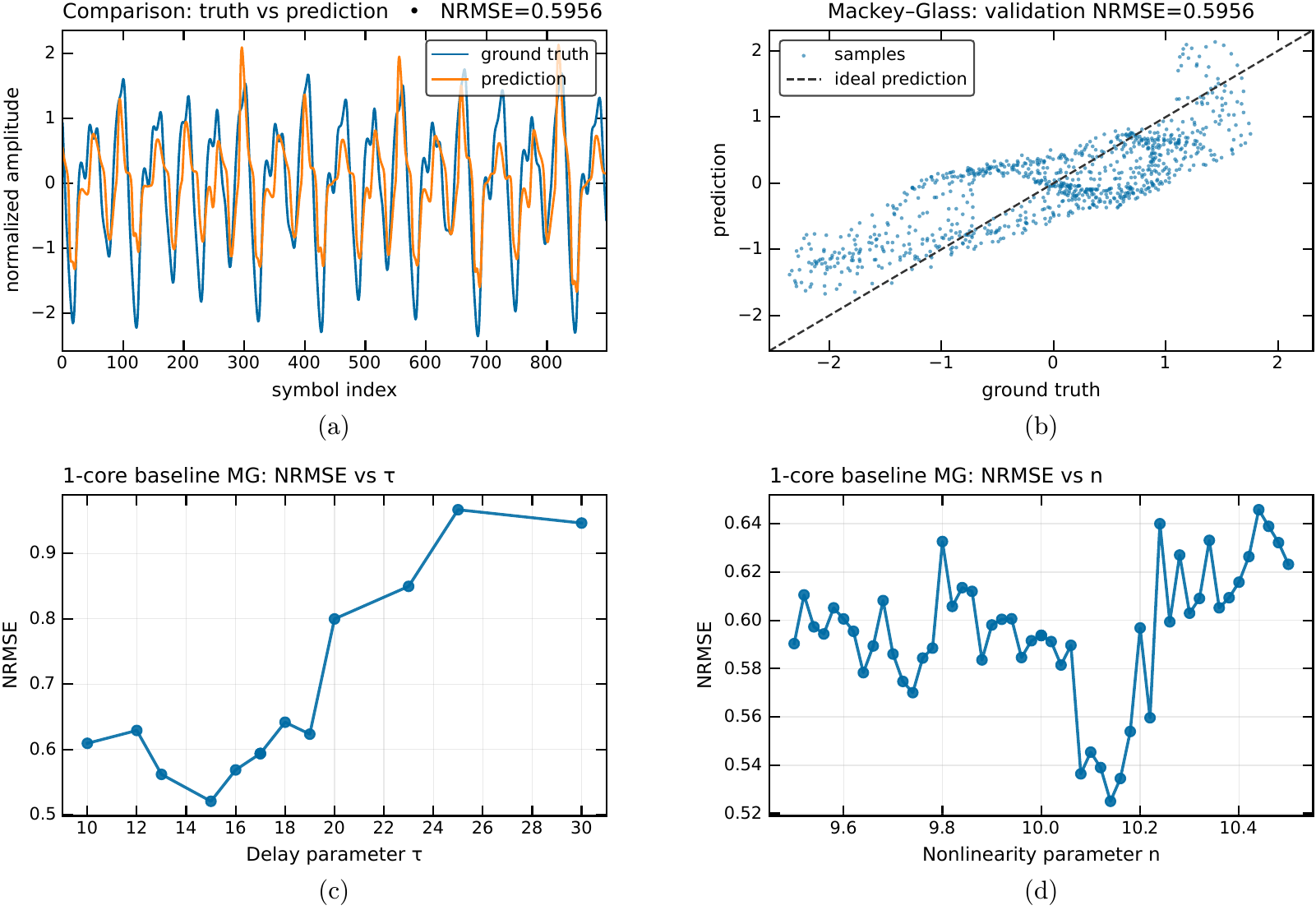}
    \caption{Results for the 1-core fiber with a temporal mask: (a) representative time-series prediction on the validation subset; (b) predicted values versus the ground truth; (c) prediction-error dependence on $\tau$ for $n=10$; (d) prediction-error dependence on $n$ for $\tau=17$.}
    \label{fig:1core_baseline_panel}
\end{figure}

Figure~\ref{fig:1core_baseline_panel}(a) shows a representative prediction result for the single-core baseline, with a validation NRMSE of 0.5956. The predicted signal follows the coarse oscillatory structure of the Mackey--Glass sequence, but the predicted values remain concentrated near the center of the target range. In particular, many local maxima and minima are substantially underestimated, indicating that the single-core reservoir does not reproduce the detailed one-step-ahead dynamics accurately.
Figure~\ref{fig:1core_baseline_panel}(b) provides a statistical view of the same limitation. Instead of forming a narrow distribution along the ideal prediction line, the predicted samples form a broad nonlinear band concentrated toward the central amplitude range. This shows a systematic underestimation of extreme target values rather than only isolated prediction failures. The present calculations do not identify a single physical cause of this behavior.

Panels~(c) and~(d) of Fig.~\ref{fig:1core_baseline_panel} show the dependence of the single-core prediction error on the Mackey--Glass equation parameters $\tau$ and $n$. The validation NRMSE remains above 0.5 for all considered parameter values, demonstrating that the baseline is consistently inaccurate throughout the explored parameter range. The dependence on $\tau$ is particularly pronounced: after reaching its lowest value near $\tau=15$, the error increases strongly for larger delay values and approaches unity for $\tau=25$--$30$. The dependence on $n$ is less systematic, but the error still fluctuates within a high range of approximately 0.52--0.65.

These results show that, for the present single-core active-fiber implementation and the explored operating conditions, temporal masking, delayed feedback and gain saturation do not generate sufficiently informative intensity features for accurate Mackey--Glass prediction. This conclusion is specific to the considered baseline and should not be interpreted as a general limitation of single-node delay reservoirs, whose performance depends strongly on the internal dynamics, masking and operating point~\cite{Appeltant2011InformationSystem}.

\subsection{Mackey--Glass: multicore reservoirs}

\subsubsection{Equal temporal masks for each core}

We next considered a 7-core hexagonal MCF in which the same temporal mask was applied to all cores. In contrast to the single-core baseline, the reservoir state now contains spatially resolved intensity samples from several coupled optical channels. Although the temporal modulation is identical across the cores, the output features need not be equivalent, because the fields propagate through the full coupled multicore system and experience an optimized pump-controlled gain profile. This regime therefore allows us to isolate the contribution of multicore propagation while retaining conventional temporal multiplexing.

\begin{figure}[!htbp]
    \centering
    \includegraphics[width=1\linewidth]{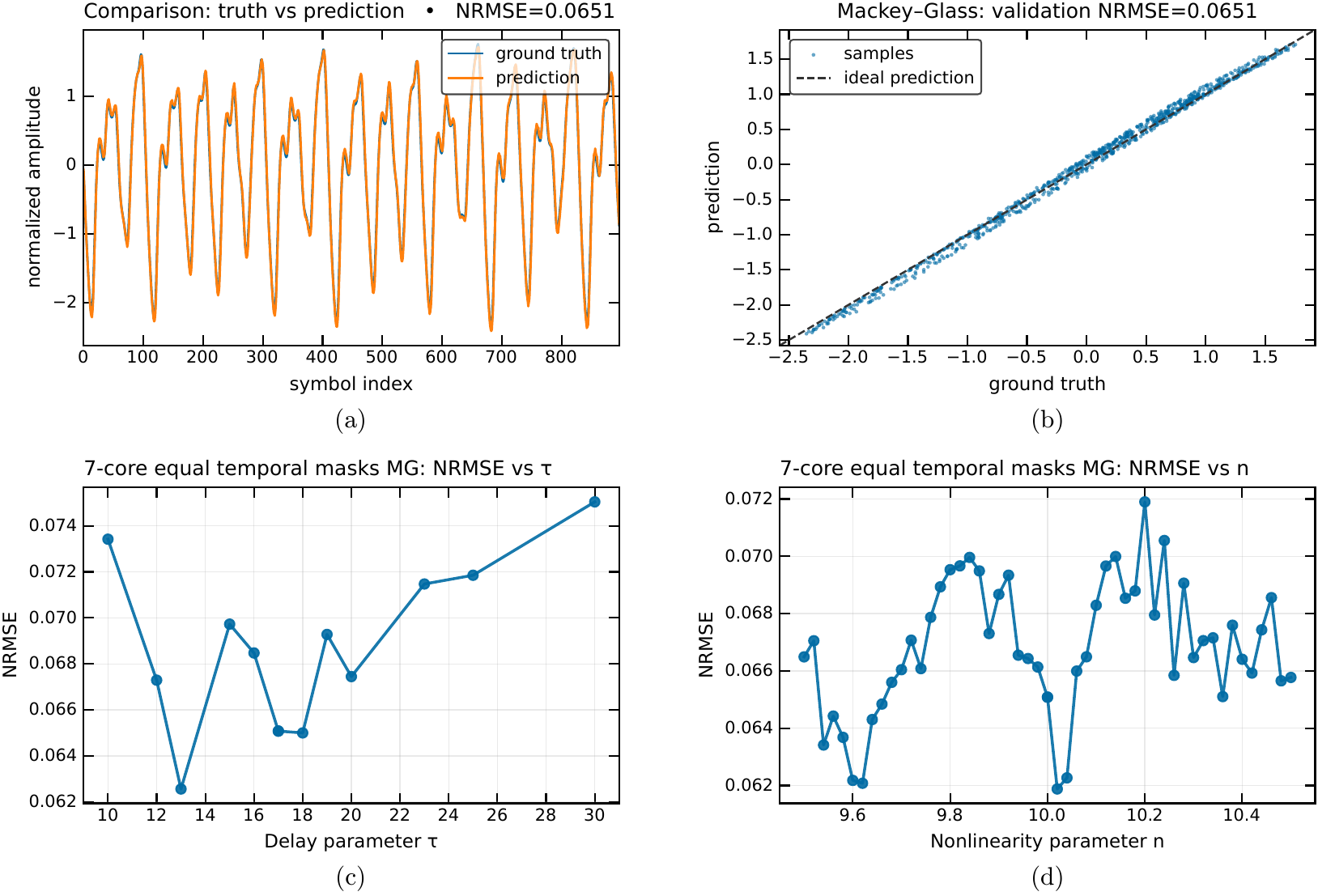}
    \caption{Results for the 7-core fiber with equal temporal masks for each core: (a) representative time-series prediction on the validation subset; (b) predicted values versus the ground truth; (c) prediction-error dependence on $\tau$ for $n=10$; (d) prediction-error dependence on $n$ for $\tau=17$.}
    \label{fig:7core_equal_masks_panel}
\end{figure}

Figure~\ref{fig:7core_equal_masks_panel}(a) shows a representative prediction result for the 7-core equal-temporal-mask regime, with a validation NRMSE of approximately 0.0651. The predicted sequence closely follows the ground truth throughout the plotted symbol window, including most local maxima, minima and rapid transitions. Compared with the single-core baseline, whose validation NRMSE is approximately 0.5956, the multicore configuration reduces the prediction error by about a factor of nine. This substantial improvement is consistent with the formation of additional useful readout features through coupled-core propagation.

Figure~\ref{fig:7core_equal_masks_panel}(b) provides a statistical view of the prediction accuracy. The predicted samples form a narrow distribution around the ideal prediction line over almost the entire target-amplitude range. Small deviations remain, particularly near the extreme amplitudes, but the systematic underestimation of extreme values observed for the single-core baseline is largely removed.

Panels~(c) and~(d) of Fig.~\ref{fig:7core_equal_masks_panel} show the dependence of the prediction error on the Mackey--Glass equation parameters $\tau$ and $n$. For all considered values, the validation NRMSE remains within a relatively narrow low-error range of approximately 0.062--0.075. The dependence on $\tau$ is nonmonotonic, with the lowest error obtained near $\tau=13$ and larger errors observed toward the boundaries of the explored interval. The dependence on $n$ also exhibits irregular fluctuations, but no systematic degradation is observed across the considered range.

These results show that the 7-core reservoir with equal temporal masks provides accurate one-step-ahead prediction not only for the reference Mackey--Glass parameters, but also across the explored variations of $\tau$ and $n$. The improvement over the single-core baseline demonstrates that multicore propagation can enrich the intensity-based reservoir state even when all cores receive the same temporal modulation.

\subsubsection{Spatial-only encoding}

Finally, we considered a 7-core MCF in the spatial-only regime at a modulation rate of 1 GHz. The intensity-readout dimension is $D=N=7$ for the 7-core case considered in detail below. In this regime, temporal memory is provided by the delayed feedback loop, while inter-core coupling, the spatial input distribution and the pump-controlled gain profile generate diverse nonlinear intensity features across the cores. This configuration therefore tests whether the spatial degrees of freedom of the MCF are sufficient for accurate prediction without serial temporal multiplexing.

\begin{figure}[!htbp]
    \centering
    \includegraphics[width=1\linewidth]{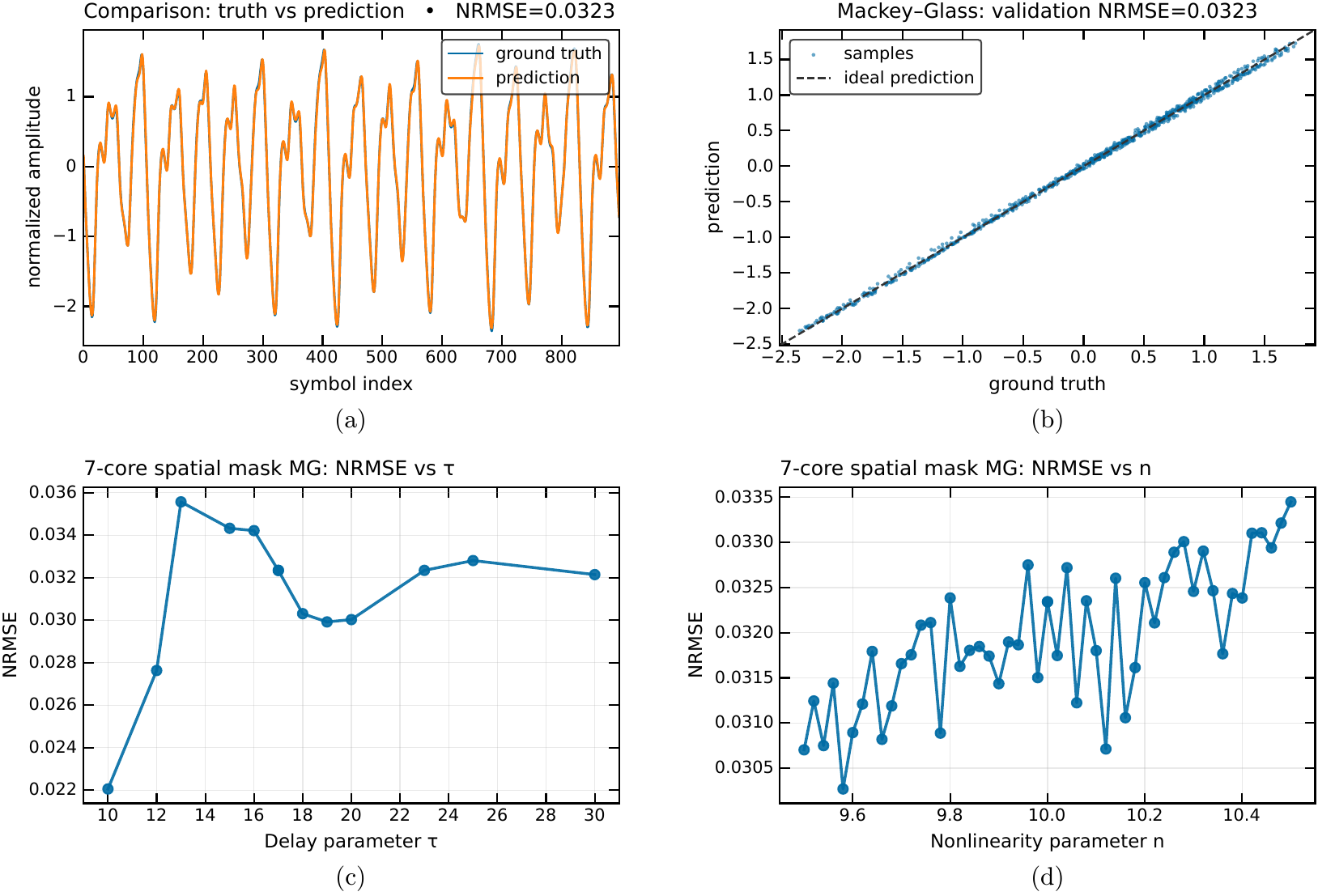}
    \caption{Results for the 7-core fiber with spatial-only encoding at 1 GHz: (a) representative time-series prediction on the validation subset; (b) predicted values versus the ground truth; (c) prediction-error dependence on $\tau$ for $n=10$; (d) prediction-error dependence on $n$ for $\tau=17$.}
    \label{fig:7core_spatial_only_panel}
\end{figure}

Figure~\ref{fig:7core_spatial_only_panel}(a) shows a representative prediction result for the spatial-only encoding regime, with a validation NRMSE of approximately 0.0323. The predicted sequence closely follows the ground truth throughout the plotted symbol window, including the local extrema and rapid transitions. Only small deviations remain, which are difficult to distinguish on the scale of the time-series plot.

Figure~\ref{fig:7core_spatial_only_panel}(b) provides a statistical view of the prediction accuracy. The predicted samples form a narrow distribution around the ideal prediction line over the full target-amplitude range. The small residual spread indicates that the seven spatially resolved intensity features contain sufficient information for accurate one-step-ahead prediction, despite the absence of a temporal mask.

Panels~(c) and~(d) of Fig.~\ref{fig:7core_spatial_only_panel} show the dependence of the prediction error on the Mackey--Glass equation parameters $\tau$ and $n$. The validation NRMSE remains low for all considered parameter values, but the sensitivity to the two parameters is different. The dependence on $\tau$ is nonmonotonic: the error varies from approximately 0.022 at $\tau=10$ to approximately 0.0355 at $\tau=13$, and then remains near 0.030--0.033 over most of the remaining interval. By contrast, the dependence on $n$ is weaker, with the NRMSE varying within a narrower range of approximately 0.0302--0.0334 and showing only a modest overall increase toward larger $n$.

These results demonstrate that, at the 1 GHz modulation rate considered here, a 7-core reservoir can accurately predict the Mackey--Glass dynamics using spatial input encoding and feedback memory. Together with the timing analysis above, the result shows why the lower modulation rate is important in the present design: it allows the feedback window to contain an MCF section long enough to produce a useful coupled-core transformation.

\section{Conclusions}

We have presented a numerical proof of concept for a reservoir-computing system based on an active multicore fiber placed inside a delayed optical feedback loop. The reservoir combines temporal memory from the feedback loop with spatially distributed optical dynamics produced by inter-core coupling, Kerr nonlinearity and pump-controlled saturable gain. The scalar input sequence is encoded through temporal and spatial masks, the optical intensities measured in the individual cores form the reservoir state, and only the final linear readout is trained. The propagation model is based on a system of linearly coupled nonlinear Schrödinger equations with saturable gain, with the fiber geometry and coupling parameters derived from a realistic multicore-fiber design.

The Mackey--Glass one-step-ahead prediction benchmark demonstrates the importance of the spatial degrees of freedom. The single-core active-fiber baseline reproduces only the coarse structure of the target sequence and has an NRMSE of approximately 0.5956, with systematic underestimation of large positive and negative excursions. By contrast, the representative 7-core reservoir with equal temporal masks reaches an NRMSE of approximately 0.0651 at 40 GHz, reducing the error by about a factor of nine. Its prediction error also remains low across the explored variations of the Mackey--Glass parameters $\tau$ and $n$. These results show that coupled-core propagation and the optimized gain profile generate useful intensity features that are not available in the single-core configuration, even when the same temporal modulation is applied to all cores.

The spatial-only regime provides a more direct demonstration of this capability. At a modulation rate of 1 GHz, the 7-core reservoir reaches an NRMSE of approximately 0.0323. The prediction remains accurate over the considered ranges of $\tau$ and $n$, indicating that spatial input encoding, multicore propagation and feedback memory can provide a sufficiently informative reservoir state without serial temporal multiplexing. With $M=1$ and $d_{\mathrm{fb}}=1$, however, the corresponding feedback window is not physically admissible at 40 GHz within the selected fiber-length range because it is shorter than the MCF group delay. The lower modulation rate therefore makes the spatial-only architecture compatible with a useful MCF length.

The main conclusion is that an active multicore fiber can serve not only as a set of parallel readout channels, but also as a tunable nonlinear transformation that enriches the reservoir state through the interaction of coupling, gain saturation and feedback. This makes multicore fibers a promising platform for reducing the reliance of delay-based reservoir computers on long temporal masks and high-rate temporal modulation. The present study is limited to a numerical model in which the main benchmark calculations neglect dispersion and are performed in the noise-free regime.

A physical implementation will also be sensitive to environmental and fabrication-induced perturbations. Drift of the optical path length changes the effective feedback phase $\Delta\phi_f$, pump-power fluctuations modify $g_{0,n}$ and $P_{\mathrm{sat},n}$, and core-to-core variations of propagation constants and coupling coefficients break the idealized symmetry of the modeled MCF. Detector noise and spatial cross-talk can further reduce the effective dimensionality of the measured state. Scaling the system will therefore require active phase stabilization, calibration of core-resolved launch and detection, monitoring of the pump distribution, and, when necessary, periodic retraining of the linear readout. The present calculations assume fixed parameters and do not quantify robustness to these perturbations. Further work should include direct comparisons of encoding strategies at the same modulation rate, evaluation on additional benchmark tasks, analysis of dispersion, ASE noise and parameter fluctuations, and experimental validation of the proposed architecture.

\section*{Acknowledgements}

I.~S. Chekhovskoy acknowledges support from ITMO University within the ITMO Fellowship Program.
The work of M.~P. Fedoruk was carried out within the state assignment of the Ministry of Science and Higher Education of the Russian Federation for the Federal Research Center for Information and Computational Technologies.

\appendix
\renewcommand{\thesection}{Appendix~\Alph{section}}

\section{MCF geometry and calculated optical parameters}
\label{app:mcf_geometry}

The core material is modeled as a silica--germania alloy. The refractive index is calculated using the composition-dependent Sellmeier model for GeO$_2$--SiO$_2$ glasses~\cite{Fleming1984Dispersion}:
\begin{equation}
\label{eq:sellmeier}
n_{\mathrm{core}}^2(\lambda,x_{\mathrm{GeO_2}})
=
1+
\sum_{j=1}^{3}
\frac{B_j(x_{\mathrm{GeO_2}})\lambda^2}
{\lambda^2-C_j(x_{\mathrm{GeO_2}})},
\end{equation}
where $x_{\mathrm{GeO_2}}$ is the germania concentration. In the simulations we used $x_{\mathrm{GeO_2}}=0.038$. The coefficients $B_j(x_{\mathrm{GeO_2}})$ and $C_j(x_{\mathrm{GeO_2}})$ are obtained from the silica--germania composition model. The cladding index is then computed from the numerical aperture as
$n_{\mathrm{clad}}
=
\sqrt{
n_{\mathrm{core}}^2-\mathrm{NA}^2
}$.
Thus, material dispersion and waveguide dispersion enter the calculation through the wavelength dependence of the refractive indices and of the guided LP$_{01}$ mode.

The propagation constant of the guided mode is treated as a function of the angular frequency,
$\beta=\beta(\omega)$, $\beta_1 = \frac{d\beta}{d\omega}$, $\beta_2 = \frac{d^2\beta}{d\omega^2}$.
In the numerical implementation, $\beta_1$ and $\beta_2$ are evaluated by finite differences with respect to the wavelength around $\lambda_0$.

The Kerr coefficient is computed from the effective area of the LP$_{01}$ mode:
\begin{equation}
\label{eq:aeff}
A_{\mathrm{eff}}
=
\frac{
\left(
\iint |F(x,y)|^2 dxdy
\right)^2
}{
\iint |F(x,y)|^4 dxdy
},
\end{equation}
and
\begin{equation}
\label{eq:gamma_from_aeff}
\gamma
=
\frac{2\pi n_2}{\lambda_0 A_{\mathrm{eff}}}.
\end{equation}

The coupling matrix $\mathbf C$ is calculated from the MCF geometry and modal overlap integrals. For two cores $n$ and $m$, the coupling coefficient is evaluated in the form~\cite{Mumtaz2013, 6253229, Fujisawa2017, RenTan2016, Agrawal}
\begin{equation}
\label{eq:coupling_coefficient}
C_{n,m}
=
\frac{k_0^2}{2\beta}
\frac{
\iint \Delta n_m^2(x,y)
F_n(x-x_n,y-y_n)
F_m(x-x_m,y-y_m)\,dx\,dy
}{
\iint F_n^2(x-x_n,y-y_n)\,dx\,dy
},
\qquad n \ne m,
\end{equation}
where $k_0$ is the vacuum wavenumber, $\beta$ is the propagation constant of the guided mode and $\Delta n_m^2$ is nonzero inside the neighboring core region. In the numerical implementation the integrals are evaluated over the finite circular cladding domain $x^2+y^2\le R_{\mathrm{cl}}^2$ for the actual hexagonal core positions, with $R_{\mathrm{cl}}=125~\mu\mathrm{m}$. In the 19-core geometry, the centers of the outermost cores are located at $2\Lambda=60~\mu\mathrm{m}$ from the fiber axis; the minimum distance from the outer-core boundary to the cladding interface is therefore $R_{\mathrm{cl}}-2\Lambda-r_c=62.05~\mu\mathrm{m}$. The finite integration boundary is thus included explicitly. However, the functions $F_n$ are still taken as isolated-core LP$_{01}$ modes and are not recomputed as finite-cladding supermodes. A full-vector eigenmode calculation of the entire MCF cross-section would be required to quantify the residual boundary-induced deformation of peripheral-core modes.

\section{Pump-to-effective-gain and gain-saturation parametrization}
\label{app:pump_gain}

The effective gain and saturation parameters are computed from the pump power assigned to each core. In terms of pump power measured in mW, the saturation power is written as
\begin{equation}
\label{eq:psat_from_pump}
P_{\mathrm{sat},n}^{(\mathrm{mW})}
=
a P_{\mathrm{pump},n}^{(\mathrm{mW})}
+
b,
\end{equation}
where
$a = 0.1292$, $b = -1.096$.
The corresponding effective gain coefficient is computed as
\begin{equation}
\label{eq:g0_from_pump}
g_{0,n}
=
\frac{
A + B P_{\mathrm{pump},n}^{(\mathrm{mW})}
}{
1 - C P_{\mathrm{pump},n}^{(\mathrm{mW})} \exp(DL)
},
\end{equation}
with
$A = -2.25\times 10^{5}~\mathrm{m^{-1}}$, $B = 2.3\times 10^{4}~\mathrm{m^{-1}mW^{-1}}$,
$C = -4.54\times 10^{3}~\mathrm{mW^{-1}}$, $D = 0.05~\mathrm{m^{-1}}$.
In the present reduced model, this coefficient is used as an effective pump-controlled gain in Eq.~\eqref{eq:SysNLSE}.
If Eq.~\eqref{eq:psat_from_pump} gives a non-positive saturation power, the corresponding core is treated as unpumped. Thus, in the reservoir simulations the optimized pump distribution controls both the gain profile and the saturation level of the active MCF.

\section{Numerical propagation method}
\label{app:numerical_propagation}

For a single propagation through the MCF section, we decompose the right-hand side of the system~\eqref{eq:SysNLSE} into the linear operator
$\mathfrak D\langle\mathbf A\rangle=i\mathbf C\,\mathbf A-\frac{g^{0}}{2}\mathbf A$, where $\mathbf A = \left( A_0, A_1, \dots, A_{N-1} \right)^T$, and the nonlinear operator $\mathfrak N\langle\mathbf A\rangle=i\gamma|\mathbf A|^{2}\mathbf A+\sigma\mathbf A+\frac{g^{0}}{2}\mathbf A$. In the more general case with dispersion, the term $-i \frac{\beta_2}{2} \frac{\partial^2}{\partial t^2} A$ is also included in $\mathfrak D$. The second-order in~$z$ scheme has the form (Strang decomposition)
\begin{equation}
\mathbf A(z+h)=
e^{\frac{h}{2}\mathfrak D}\,
e^{h\mathfrak N}\,
e^{\frac{h}{2}\mathfrak D}\mathbf A(z),
\label{eq:strang}
\end{equation}
where the matrix exponential $e^{\frac{h}{2}\mathfrak D}$ is calculated once for the current MCF configuration. In the dispersive case this operation is performed in Fourier space using Pad\'{e} approximation~\cite{ChekhovskoyEtAl_JoCP_2017, Moler2003, Higham}; in the simulations with $\beta_2=0$ it reduces to a matrix operation in the time domain.
For nonlinear half-stepping, the Madelung representation $A_{n}=\sqrt{P_{n}}\exp(i\phi_{n})$, is used, which gives closed expressions (analytical solution for the nonlinear step)
\begin{align}
E_{n}(z) &= \sqrt{\bigl[E_{n}(0)^{2}+2E_{n}(0)E_{\text{sat},n}\bigr]
e^{\,g^{0}_{n} z}+{E_{\text{sat},n}}^{2}}
-E_{\text{sat},n},
\label{eq:energy_nonlinear_step}\\
P_{n}(t,z) &= P_{n}(t,0)\,
\sqrt{\frac{E_{n}(z)\bigl[E_{n}(0)+2E_{\text{sat},n}\bigr]}
{E_{n}(0)\bigl[E_{n}(z)+2E_{\text{sat},n}\bigr]}}
\,e^{\,g^{0}_{n} z},
\label{eq:power}\\
\phi_{n}(t,z) &= \phi_{n}(t,0)+
\frac{\gamma_{n}P_{n}(t,0)}{g^{0}_{n}E_{n}(0)}
\Bigl[E_{n}(z)-E_{n}(0)-E_{\text{sat},n}
\ln\!\frac{E_{n}(z)+2E_{\text{sat},n}}{E_{n}(0)+2E_{\text{sat},n}}\Bigr],
\label{eq:phase}
\end{align}
after which the new value of the field is restored by $A_{n}=\sqrt{P_{n}}\exp(i\phi_{n})$.
The method remains non-iterative for a single MCF pass, has the second order in the coordinate $z$ and is stable for the considered operating regimes, while the computational complexity of one step is reduced to $O(N^{2}M)$, where $M$ is the size of computational grid in time. Adjacent half-steps with the linear operator can be combined due to commutativity, so the number of operations can be reduced.

The longitudinal step (by $z$) is selected using characteristic propagation lengths. For the input field $q(t)$ in the central core, we estimate
\begin{equation}
\label{eq:dispersion_length}
L_D =
\frac{2E}{|\beta_2|\int\limits_{-T/2}^{T/2}\left|\frac{\partial q}{\partial t}\right|^2dt},
\qquad
E =
\int\limits_{-T/2}^{T/2}|q|^2dt,
\end{equation}
and
\begin{equation}
\label{eq:characteristic_lengths}
L_{NL} =
\frac{E}{\gamma\int\limits_{-T/2}^{T/2}|q|^4dt},
\qquad
L_c = \frac{\pi}{2\max\limits_{n\ne m}|C_{n,m}|},
\qquad
L_g = \frac{1}{\max_n g_{0,n}}.
\end{equation}
In the simulations with $\beta_2=0$, one has $L_D = \infty$, and the spatial step is controlled by the nonlinear, coupling and gain lengths. We set
\begin{equation}
\label{eq:z_steps}
L_{\min} = \min(L_D,L_{NL},L_c,L_g),
\qquad
N_z =
\max\left(
\operatorname{round}\left(N_{\mathrm{pp}}\frac{L}{L_{\min}}\right),
1
\right),
\end{equation}
where $N_{\mathrm{pp}}$ is the number of longitudinal steps per characteristic length. In the reported simulations we used $N_{\mathrm{pp}}=20$ unless stated otherwise. This value was chosen as a practical resolution margin for the second-order split-step scheme while keeping the feedback-loop simulations computationally feasible.

\section{Feedback-loop timing}
\label{app:feedback_timing}

The total feedback delay consists of two parts: the group-delay time spent by the optical field inside the MCF section and the delay accumulated in the external part of the feedback loop. In the retarded-time propagation model, the MCF group delay does not appear in the propagation equation itself, but it must still be included when the physical round-trip time of the feedback loop is specified. Therefore, for a selected computational window of duration $T_{\mathrm{win}}$, we set
\begin{equation}
\label{eq:feedback_total_time}
T_f
=
\beta_1 L
+
T_{\mathrm{free}}
=
T_{\mathrm{win}},
\end{equation}
or equivalently
\begin{equation}
\label{eq:feedback_free_space_time}
T_{\mathrm{free}}
=
T_{\mathrm{win}}
-
\beta_1 L.
\end{equation}
Here $T_{\mathrm{free}}$ denotes the delay that must be provided by the external part of the loop, including free-space propagation and other optical elements outside the MCF. Only parameter sets with $T_{\mathrm{free}}>0$ are physically admissible in this model, because otherwise the selected computational window would be shorter than the propagation time through the MCF section itself.

\section{ASE-noise model and effective OSNR}
\label{app:ase_osnr}

After the optical signal passes through the MCF, ASE noise can be added as an integral noise contribution from the active fiber section. This noise is added after the MCF propagation step; therefore, the noisy field is used both as the output field and as the field entering the next feedback round trip. In the main simulations the noise option is switched off. When ASE is enabled, the noise factor $\text{nf}$ is used as the main parameter of the amplifying medium. The effective gain in the $n$-th core is estimated as
\begin{equation}
G_n = \exp(g^0_n L).
\label{eq:noise_gain}
\end{equation}
For finite gain, the spontaneous emission factor is computed from
\begin{equation}
NF_{\text{lin}}
=
\frac{2n_{\text{sp},n}(G_n-1)+1}{G_n},
\qquad
n_{\text{sp},n}
=
\frac{NF_{\text{lin}}G_n-1}{2(G_n-1)}.
\label{eq:noise_factor}
\end{equation}
The ASE spectral density and the discrete noise power are then taken as
\begin{equation}
S_{\text{ASE},n}
=
n_{\text{pol}} n_{\text{sp},n} h\nu_s (G_n-1),
\qquad
P_{\text{ASE},n}
=
S_{\text{ASE},n}B_{\text{ASE}},
\label{eq:noise_power}
\end{equation}
where $n_{\text{pol}}=1$ is the number of polarizations included in the scalar noise model, $h\nu_s$ is the photon energy, and $B_{\text{ASE}} = 12.5$ GHz is the effective optical noise bandwidth used in the simulation. The resulting noise is modeled as additive unbiased complex white Gaussian noise with
\begin{equation}
\mathbb{E}\left[|\eta_n(t)|^2\right]=P_{\text{ASE},n}.
\label{eq:noise_variance}
\end{equation}
Thus, for a noisy run the field entering the detector and the feedback loop is
\begin{equation}
A_{n,\text{out}}^{\text{noisy}}(t)
=
A_{n,\text{out}}(t)+\eta_n(t).
\label{eq:noisy_output}
\end{equation}

In addition to this nominal ASE estimate, we compute an effective output OSNR directly from two numerical runs. The first run is performed with ASE noise enabled, and the second run is repeated with the same reservoir parameters but with ASE noise disabled. The effective noise contribution is obtained as the difference between the noisy and noiseless output fields on the same time interval:
\begin{equation}
\text{OSNR}_{\text{eff}}
=
10\log_{10}
\frac{
\sum_{n,t}|A_{n,\text{out}}^{\text{clean}}(t)|^2
}{
\sum_{n,t}|A_{n,\text{out}}^{\text{noisy}}(t)-A_{n,\text{out}}^{\text{clean}}(t)|^2
}.
\label{eq:effective_osnr}
\end{equation}
This quantity should be interpreted as the effective output OSNR of the simulated reservoir, because it includes the actual nonlinear propagation, gain saturation and feedback dynamics of the full model.

\bibliographystyle{unsrt}
\bibliography{references,references2}

\end{document}